\documentclass[aip,jap,reprint]{revtex4-1}
\usepackage{graphicx}

\begin{document}
\title{Optical Study of GaAs quantum dots embedded into AlGaAs nanowires.}

\author{V.N.~Kats}
\author{V.P.~Kochereshko}
\author{A.V.~Platonov}
\author{T.V.~Chizhova}
\author{G.E.~Cirlin}
    \altaffiliation[Also at ]{St. Petersburg Academic University of the RAS, Khlopina 8/3, 195220 St. Petersburg, Russia }
\author{A.D.~Bouravleuv}
    \altaffiliation[Also at ]{St. Petersburg Academic University of the RAS, Khlopina 8/3, 195220 St. Petersburg, Russia }
\author{Yu.B.~Samsonenko}
    \altaffiliation[Also at ]{St. Petersburg Academic University of the RAS, Khlopina 8/3, 195220 St. Petersburg, Russia }
\author{I.P.~Soshnikov}
    \altaffiliation[Also at ]{St. Petersburg Academic University of the RAS, Khlopina 8/3, 195220 St. Petersburg, Russia }
  \affiliation{Ioffe Physical-Technical Institute of the RAS, Politekhnicheskaya 26, 194021 St. Petersburg, Russia}
\author{E.V.~Ubyivovk}
   \affiliation{St. Petersburg State University, Physical Department, Ulyanovskaya 1, Peterhof, 198504 St. Petersburg, Russia}
\author{J.~Bleuse}
\author{H.~Mariette}
\affiliation{CEA-CNRS group ''Nanophysique et Semiconducteurs'',
CEA, INAC, SP2M, and Institut N\'{e}el, 17 rue des Martyrs, F-38054
Grenoble, France}

\date{\today}

\begin{abstract} We report on the photoluminescence characterization of
GaAs quantum dots embedded into AlGaAs nano-wires. Time integrated and time
resolved photoluminescence measurements from both an array and a single
quantum dot/nano-wire are reported. The influence of the diameter sizes
distribution is evidenced in the optical spectroscopy data together with the
presence of various crystalline phases in the AlGaAs nanowires.
\end{abstract}

\pacs{61.46.Km, 78.55.Cr, 78.47.jd, 68.37.Hk, 68.37.Lp}% PACS, the Physics and Astronomy
                             % Classification Scheme.
\keywords{AlGaAs nanowires, quantum dots, micro-photoluminescence}%Use showkeys class option if keyword
                              %display desired
\maketitle

\section*{Introduction }

Free-standing semiconductor nanowires (NWs) are promising one-dimensional
nanostructures for applications in nanoelectronics [\cite{Bryl1}], and
nanophotonics [\cite{Grad2}], as well as for fundamental research [\cite{Schub3,Seif4,Dubr5,Harm6,Plan7,Dubr8}].
 The technology of fabrication of such NWs is developing
intensively nowadays [\cite{Glas9,Pers10,Gud11,Bj12,Cui13,Pat14}].
Recently, an opportunity has been demonstrated to embed one or
several quantum dots into NW. Modern epitaxial techniques enable one
in principle to control the sizes and positions of NW at the
nanometer scale, which is very attractive for applications. The
important task for today in this area is to obtain highly homogenous
ensembles of NWs that have a small distribution of sizes and
well-determined spatial positions, in order to use them in
nanoelectronics. Despite very impressive progresses in nanowire
fabrication, many growth-related aspects in this field are not
completely understood so far. In particular, it is very important to
determine the growth procedure (temperature, flow rates etc) that
allows one to produce homogenous NWs ensembles.

\section{Experiment }

In these studies we used AlGaAs nanowires (NWs) with embedded GaAs quantum
dots (QDs). AlGaAs NWs were grown on GaAs (111)B semi-insulating substrate
using EP1203 MBE system equipped with solid sources supplying Ga and Al
atoms, and an As effusion cell to produce tetramers. The substrate surface
was first deoxidized at 630\r{ }C, then a 100~nm thick GaAs buffer layer was
grown at 600\r{ }C to achieve an atomically-flat surface. A total amount of
Au equivalent to 1 nm layer was deposited without V-group flux at 550\r{ }C
using an Au effusion cell installed directly into the III-V growth chamber
followed by a 1 min waiting time to achieve better homogeneity of the
droplets. The substrate temperature was then set to the desired value for
the growth (550\r{ }C or 580\r{ }C). This procedure induces the formation of
droplets containing Au alloyed with the substrate constituents. The typical
distribution of droplet sizes ranges between 40~nm and 50~nm. The NW growth
was initiated by opening simultaneously the Al, Ga and As sources. For the
samples the nominal growth rates, i.e. the growth rate on a clean and
Au-free surface, were fixed at 1 monolayer (ML)/s for GaAs and AlAs 0.4
ML/s. The resulting Al$_{x}$Ga$_{1-x}$As growth rate was 1.4 ML/s
corresponding to an Al content in the solid solution of $x$ = 0.285 for a
planar layer. As the Al content in the NW and planar layer could be not the
same, the NW composition was measured especially by Raman spectroscopy. We
have found that in our NWs $x$ = 0.24 -- 0.26.

For AlGaAs/GaAs/AlGaAs quantum dot formation, the growth started with 15~min
of AlGaAs, and then the Al source was closed for 5~sec in order to form a
GaAs slice in each NW. Then, the growth of AlGaAs was performed again to
produce a core -- shell structure. The growth was completed with 2~min
deposition of GaAs at 530\r{ }C to avoid possible oxidation of AlGaAs shell
layer. We did not perform any growth interruption at the heterointerfaces.
The diameter of NWs in our samples was 30-60~nm. Scanning electron
microscopy (SEM) and transmission electron microscopy (TEM)
characterizations were performed before the optical measurements.

The TEM image of the single AlGaAs NW with a slice of GaAs is shown
in Fig.~1. The dark area on the image in the center of the single NW
corresponds to the single GaAs QD. It is clearly seen that the
interfaces along the axis of growth and along the radius are rather
sharp. The GaAs QD looks like a disk surrounded by wide-gap
semiconductor AlGaAs. The diameter of the NW in our samples was
25-50~nm, the length $\propto $ 500~nm, and the thickness of the QD
was of several mono-atomic layers i.e. $\propto $~2-5~nm.

\begin{figure}[tb]
\centerline{\includegraphics[width=0.8\linewidth]{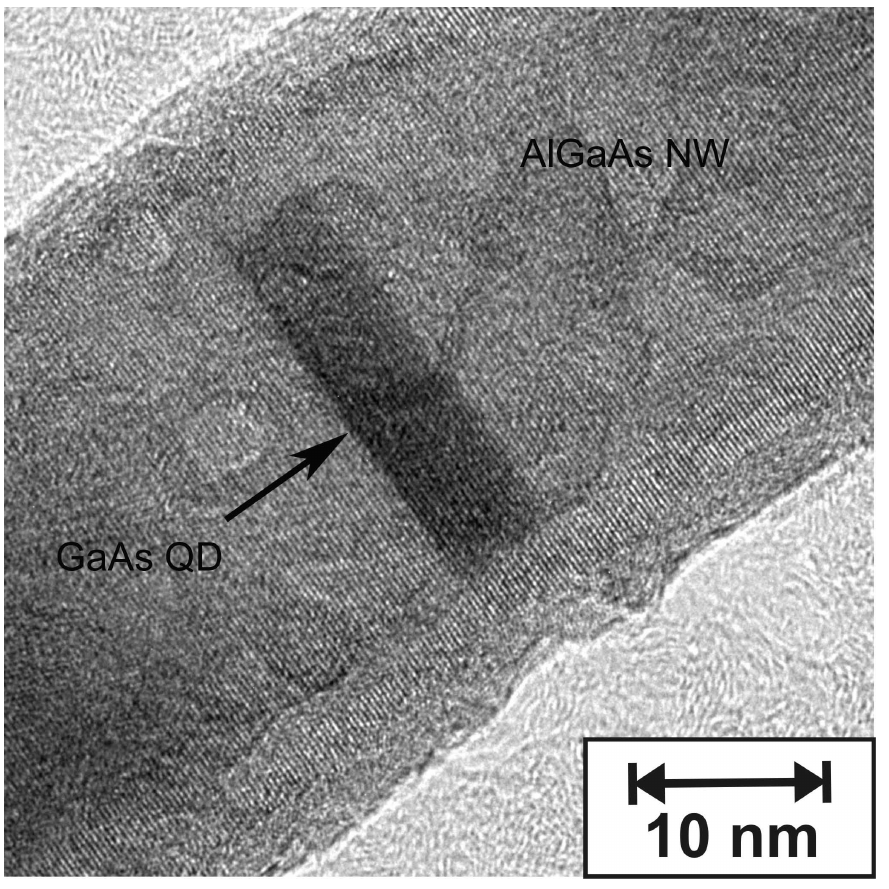}}
\caption{TEM image of the AlGaAs nanowire with embedded GaAs QD.}
\label{fig1}
\end{figure}

A systematic steady state and time resolved photoluminescence (PL)
study was performed on a set of samples in the temperature range
from 5~K up to 250~K and for various optical excitation densities.
The samples were excited by either second harmonic of Nd:YAG laser
or doubled Ti:Sa laser with  100 fs pulse duration. Jobin-Yvon 30~cm
monochromator and avalanche photodiode were used for registration.

\section{Results }

Figure.2a shows an SEM image of the sample grown at "high"
temperature (580\r{ }C). It is seen that the nano-wires are strongly
inhomogeneous in diameters and positions on the substrate.

The diameter dispersion ranges between 100~{\AA} and 500~{\AA}, and
can even change along a given wire. Most of the wires are
perpendicular to the substrate, but some of them are lying
horizontally, revealing also the orientations inhomogeneity and some
of them are cut from the substrate at all. It is clear that this
sample is strongly inhomogeneous.

Photoluminescence spectrum from this type of sample is shown on
Fig.2b The spectrum is very broad and spreads from 1.65~eV to
1.95~eV without any reproducible structure. This emission band is a
mixture of recombination coming from NWs and QDs embedded into them,
without any possibility to separate both contributions.

Instead, for samples grown at lower temperature (550\r{ }C), a much
better homogeneity is obtained for both the diameters and the length
of the NWs (see Fig.3a). For example, we can estimate from this figure
that the distribution of the NW diameters does not exceed 15~{\%}.
Additionally, the diameter does not change along the wire.

\begin{figure}[t]
\centerline{\includegraphics[width=0.8\linewidth]{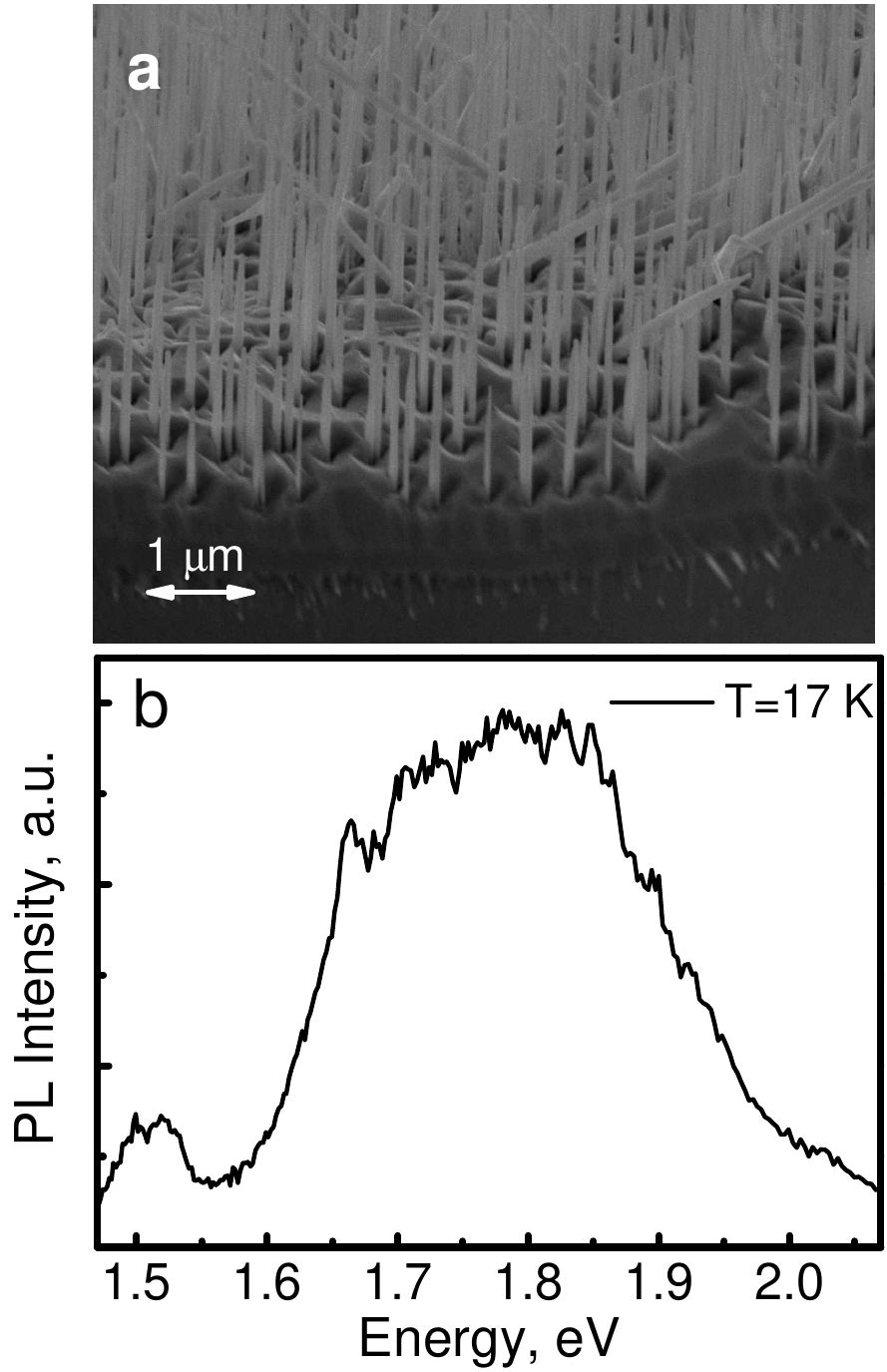}}
\caption{a$)$. SEM image of the sample grown at  580\r{ }C. \\
b$)$. Photoluminescence spectrum taken from the given sample. }
\label{fig2}
\end{figure}

The PL spectrum taken from this structure at temperature of 10~K is
presented in Fig.3b by solid line. Two intense lines at energies
1.73~eV and 1.87~eV are dominate in the spectrum. These lines are
attributed to carrier recombination respectively in the quantum dots
and in the nanowires.

Indeed, the nano-wire material Al$_{x}$Ga$_{1-x}$As, $x\approx
0.25-0.3$ has bandgap $E_g^{AlGaAs} $=1.8 - 1.9~eV, and the bandgap
of the quantum dot material is $E_g^{GaAs} $=1.519~eV at helium
temperatures. Consequently, the PL line of quantum dots can has
energies from $E_g^{GaAs} $=1.519~eV, in very large dots to
$E_g^{AlGaAs} $=1.8~eV, in very small dots. From the other hand the
PL line of NW can not be lower in energy than the
Al$_{0.25}$Ga$_{0.75}$As bandgap, i.e. can not be below
$E_g^{AlGaAs} $=1.8~eV. By contrast with the results shown on figure
2b, the width of these PL lines reveals a much narrower distribution
of NW diameters, which do not exceed 15{\%}.

\begin{figure}[tb]
\centerline{\includegraphics[width=0.8\linewidth]{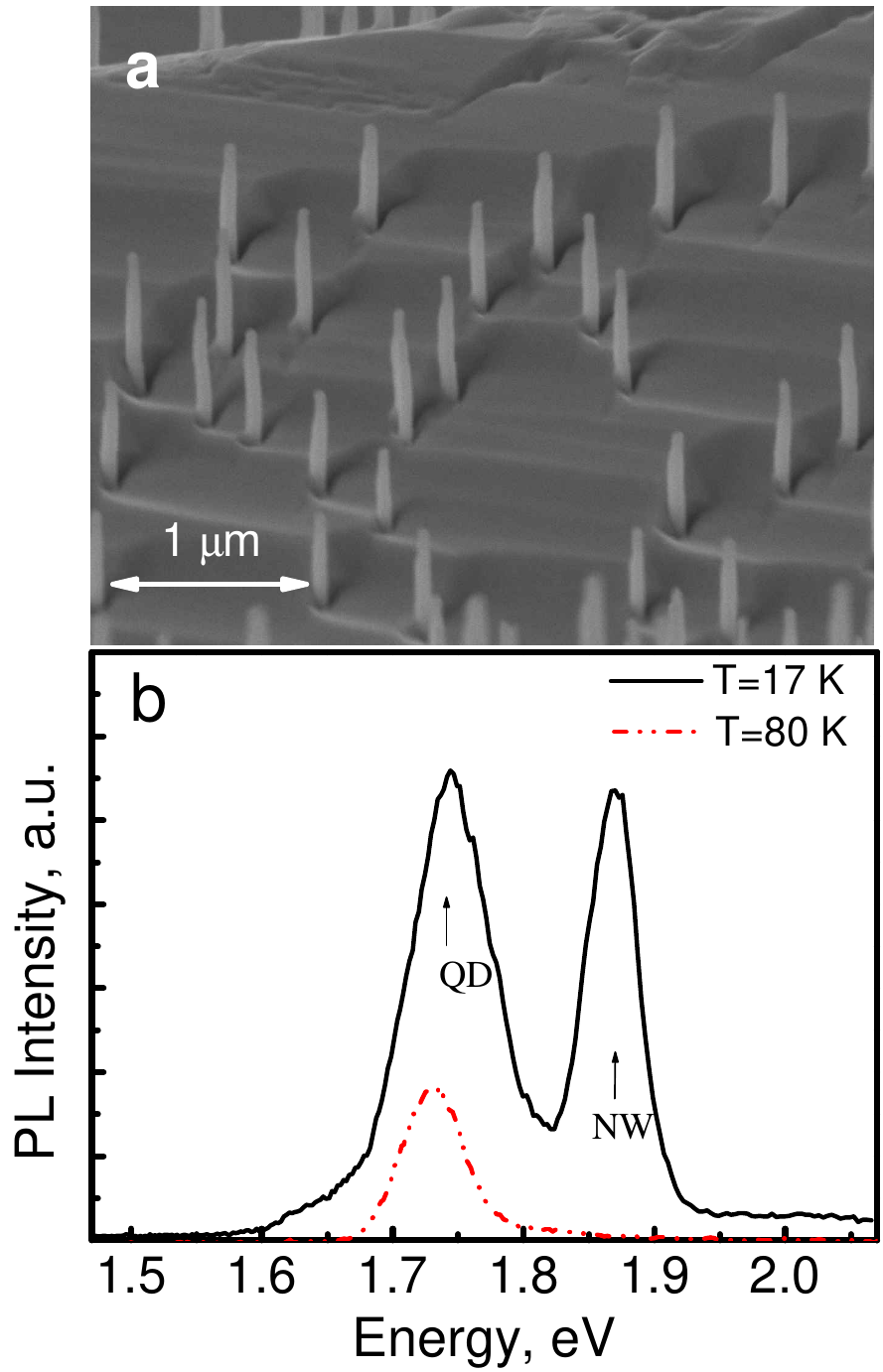}}
\caption{a$)$. SEM image of the  sample grown at 550\r{ }C. \\
b$)$. Photoluminescence spectra taken at two temperatures for the given sample . } \label{fig3}
\end{figure}

The difference in the nature of these two lines shown in figure 3b
is confirmed by the temperature dependence of the emission spectrum.
The PL spectrum taken at 77K is presented in Fig.3b by the dashed
line: the intensity of the nano-wire line completely disappears with
temperature, whereas the one of the QD line drops down by only a
factor three. This evidences the strong localization of the carriers
at the origin of the QD line, by contrast to the one in the
recombination of the line at higher energy (1.87~eV) which can
diffuse along the wire axis and be capture by the QD.

Comparing the results shown in Fig.2 and Fig.3 we can conclude that
it is possible to obtain a homogenous distribution of NW diameters
by adjusting the growth conditions.

\begin{figure}[tb]
\centerline{\includegraphics[width=0.8\linewidth]{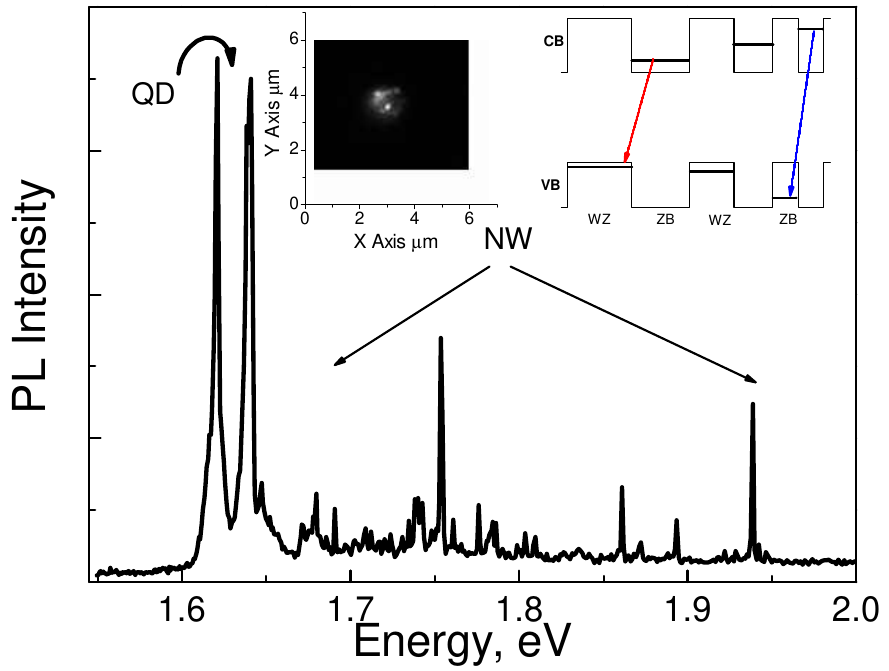}}
\caption{Micro-Photoluminescence spectrum taken at T= 6~K from the GaAs/AlGaAs QD/NW structure.\\
Left inset: Spatially resolved PL of the area under study.\\
Right inset: The scheme of the effective type-II "superlattice" formation. } \label{fig4}
\end{figure}

Having optimized the growth process to obtain a homogeneous
distribution of NW diameters, we performed an optical study (PL
spectra and time resolved spectroscopy) on a single QD and single
NW. For this purpose a strongly increased image of the sample
surface was projected into the slit of the spectrometer. As a
result, only the signal coming from a very small number of NWs is
collected. The spatial resolution of the objective was
$\propto$~1.5~micron, and the average distance between nano-wires
was about $\propto$~0.8~microns. Consequently the photodetector can
collect a signal coming from several nano-wires contained quantum
dots of little different sizes. Spatially resolved PL image  is
shown in the right  insert to Fig.4a. In the present case, the emission
spectrum shown on Fig.4a, corresponds to three NWs having a diameter
nearly 50~A bigger than the one of the sample presented in Fig.3.
There are two groups of lines in this spectrum: three relatively
wide lines in the energy range 1.6 -- 1.65~eV and many irregular
distributed narrow lines at higher energy (1.75 -- 1.95~eV).

\begin{figure}[tb]
\centerline{\includegraphics[width=0.8\linewidth]{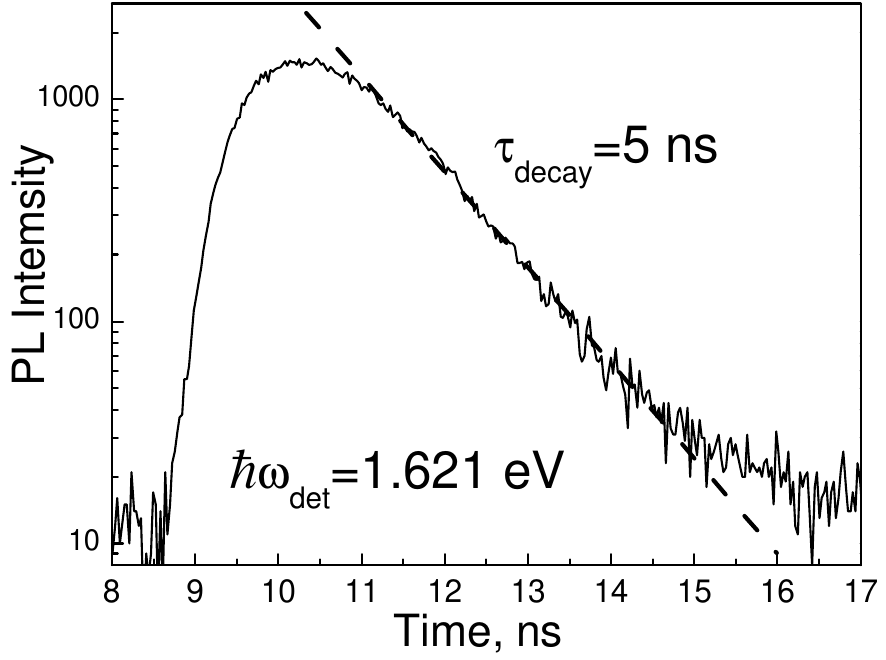}}
\caption{Time dependence of the PL maximum for the lowest QD peak on the figure 4.} \label{fig5}
\end{figure}

The line at 1.62~eV and the doublet at 1.64~eV are attributed to the
PL from the three single quantum dots embedded in the NWs. Such
interpretation is confirmed by time resolve PL measurements, as
shown in Fig.5. The decay time for all these three lines was found
to be the same and equal to 5~nsec. This is a little bit longer than
the value obtained for the decay time of the exciton recombination
in self-assembled (Stranski - Krastanov) quantum dots, but is
remarkably shorter than the decay time obtained for all other lines
(20~nsec) in this spectrum.

The full width at half maximum (FWHM) of the emission line of a
single quantum dot was found to be relatively large (typically
0.2~meV). Such linewidth is not lifetime limited; it implies instead
that the linewidth is not an intrinsic property of genuine quantum
dots, but a consequence of their local environment, which can be
explained by fluctuating Stark shift caused by photo-excited charged
carriers. Such results were already reported \cite{Kur18} for
lattice matched quantum dots like GaAs/AlGaAs by using a droplet
epitaxy method.

The origin of the linewidth could be due also to the centre of mass
quantization of the exciton in the transverse direction of the NW:
indeed, quantum dots in our samples have a disk shape, with
thickness of about 20 -- 50~{\AA} and diameter of 200 -- 500~{\AA}.
The electron quantization energy in such disks is tens and even
hundreds meV for quantization along the NW axis, but nearly one
hundred times less for "transverse" quantization. As a result the
energy levels of the transverse quantization will concentrate in the
vicinity of the main level of the longitudinal quantization. The
lines of allowed optical transitions in this case will look as an
array of narrow lines of smaller intensity concentrating around the
main line.

At high enough optical excitation all these closely situated levels are
populated by carriers. This leads to the large observed line-width and to
the increasing of the PL decay time up to 5~nsec. Unfortunately it was not
possible in our experiments to decrease the excitation density at a value
small enough without loosing the signal detected by the system. However, by
increasing the photoexcitation power, these lines become broader.

Then all the optical data for the low energy lines are consistent with the
emission coming from highly populated quantum dots in the NWs.

Concerning the array of irregular distributed lines at higher
energies (1.75~eV -- 1.95~eV), there are attributed to carriers
recombination in single nano-wires. As it was mention before, the
effective bandgap of the nano-wire can not be smaller the bandgap of
bulk Al$_{\mbox{õ}}$Ga$_{(1-\mbox{õ})}$As -- the material in which
the nano-wire made. However, in our spectrum (Fig. 4a) the line at
1.75~eV is at about 50~meV less than the bandgap energy of the
AlGaAs with x = 25{\%}. The decay time of the PL for this line is
long - 20~nsec, and it disappears at temperature of 77~K which is
typical for a PL signal coming from carriers in the NW.

We suppose that the appearance of these lines are connected with
spontaneous formation of mix phase (wurtzite)-(zinc blend).

The observation of such lines in the range 1.75 -- 1.95~eV with an
irregular distribution can be explained taking into account the
effect of the crystalline phase transition which is known to exist
in such nano-wires [\cite{Sht15,Dub16,Spir17}]. Namely in the AlGaAs
nano-wire the crystalline structure can switch spontaneously from
the zinc-blende (ZB) to the hexagonal wurtzite (WZ) structure. Since
there is a difference for the band gap for ZB and WZ AlGaAs, the
alternation in crystalline structure leads to the formation of an
effective type-II "superlattice" along the growth direction (see the left
insert of Fig4a) inducing a quantization of the carriers along the
growth axis. Consequently, the spectrum of the carriers is
transformed into a quasi-zero-dimensional one rather than a
one-dimensional one specific to the homogenous wire. It is clear
that the quantization energy depends on the thickness of the layers
with the same crystalline structure. Since the thickness was an
uncontrolled parameter in our setup we obtain a random distribution
of the carrier energy states as illustrated on the left insert of
Fig.~4a).

The long decay time, typically 20~nsec) for all these lines in the range
from 1.75~eV -- 1.95~eV confirms such interpretation, namely these lines are
related to the spatially indirect recombination in type-II structures.

\section{Conclusion}

Photoluminescence spectra from the QD embedded into NW have been measured.
It has been shown that by adjusting growth procedure one can obtain a
homogenous distribution for the QD/NW sizes. The micro-PL spectroscopy
reveals the carriers recombination both into the QDs and into the NWs. For
the first one FWHM and decay times values appear to be larger than the one
obtained in self-assembled QDs which could be due to not only some spectral
diffusion but also to some influence of the transverse quantization. As far
as the second are concerned there are attributed to recombination between
electrons in the ZB AlGaAs layers and holes in the WZ AlGaAs layers. The
long decays obtained for these latter transitions strongly support such
interpretation on the formation of type II superlattice between ZB and WZ
phases.

\begin{acknowledgments} This work was supported in part by
grants of Presidium RAS, the Russian Federal Agency for Science and
Innovation (Contract No. 02.740.11.0383), different RFBR grants and the
program of cooperation between RFBR and CNRS.
\end{acknowledgments}

\providecommand{\noopsort}[1]{}\providecommand{\singleletter}[1]{#1}%
%

%\bibliography{JAP_HM_FIN}

\begin{thebibliography}{18}%
\makeatletter
\providecommand \@ifxundefined [1]{%
 \@ifx{#1\undefined}
}%
\providecommand \@ifnum [1]{%
 \ifnum #1\expandafter \@firstoftwo
 \else \expandafter \@secondoftwo
 \fi
}%
\providecommand \@ifx [1]{%
 \ifx #1\expandafter \@firstoftwo
 \else \expandafter \@secondoftwo
 \fi
}%
\providecommand \natexlab [1]{#1}%
\providecommand \enquote  [1]{``#1''}%
\providecommand \bibnamefont  [1]{#1}%
\providecommand \bibfnamefont [1]{#1}%
\providecommand \citenamefont [1]{#1}%
\providecommand \href@noop [0]{\@secondoftwo}%
\providecommand \href [0]{\begingroup \@sanitize@url \@href}%
\providecommand \@href[1]{\@@startlink{#1}\@@href}%
\providecommand \@@href[1]{\endgroup#1\@@endlink}%
\providecommand \@sanitize@url [0]{\catcode `\\12\catcode `\$12\catcode
  `\&12\catcode `\#12\catcode `\^12\catcode `\_12\catcode `\%12\relax}%
\providecommand \@@startlink[1]{}%
\providecommand \@@endlink[0]{}%
\providecommand \url  [0]{\begingroup\@sanitize@url \@url }%
\providecommand \@url [1]{\endgroup\@href {#1}{\urlprefix }}%
\providecommand \urlprefix  [0]{URL }%
\providecommand \Eprint [0]{\href }%
\providecommand \doibase [0]{http://dx.doi.org/}%
\providecommand \selectlanguage [0]{\@gobble}%
\providecommand \bibinfo  [0]{\@secondoftwo}%
\providecommand \bibfield  [0]{\@secondoftwo}%
\providecommand \translation [1]{[#1]}%
\providecommand \BibitemOpen [0]{}%
\providecommand \bibitemStop [0]{}%
\providecommand \bibitemNoStop [0]{.\EOS\space}%
\providecommand \EOS [0]{\spacefactor3000\relax}%
\providecommand \BibitemShut  [1]{\csname bibitem#1\endcsname}%
\let\auto@bib@innerbib\@empty
%</preamble>
\bibitem [{\citenamefont {Bryllert}\ \emph {et~al.}(2006)\citenamefont
  {Bryllert}, \citenamefont {Wernersson}, \citenamefont {Lowgren},\ and\
  \citenamefont {Samuelson}}]{Bryl1}%
  \BibitemOpen
  \bibfield  {author} {\bibinfo {author} {\bibfnamefont {T.}~\bibnamefont
  {Bryllert}}, \bibinfo {author} {\bibfnamefont {L.~E.}\ \bibnamefont
  {Wernersson}}, \bibinfo {author} {\bibfnamefont {T.}~\bibnamefont {Lowgren}},
  \ and\ \bibinfo {author} {\bibfnamefont {L.}~\bibnamefont {Samuelson}},\
  }\href@noop {} {\bibfield  {journal} {\bibinfo  {journal} {Nanotechnology}\
  }\textbf {\bibinfo {volume} {17}},\ \bibinfo {pages} {S227} (\bibinfo {year}
  {2006})}\BibitemShut {NoStop}%
\bibitem [{\citenamefont {Gradecak}\ \emph {et~al.}(2005)\citenamefont
  {Gradecak}, \citenamefont {Qian}, \citenamefont {Li}, \citenamefont {Park},\
  and\ \citenamefont {Lieber}}]{Grad2}%
  \BibitemOpen
  \bibfield  {author} {\bibinfo {author} {\bibfnamefont {S.}~\bibnamefont
  {Gradecak}}, \bibinfo {author} {\bibfnamefont {F.}~\bibnamefont {Qian}},
  \bibinfo {author} {\bibfnamefont {Y.}~\bibnamefont {Li}}, \bibinfo {author}
  {\bibfnamefont {H.~G.}\ \bibnamefont {Park}}, \ and\ \bibinfo {author}
  {\bibfnamefont {C.~M.}\ \bibnamefont {Lieber}},\ }\href@noop {} {\bibfield
  {journal} {\bibinfo  {journal} {Appl. Phys. Lett.}\ }\textbf {\bibinfo
  {volume} {87}},\ \bibinfo {pages} {173111} (\bibinfo {year}
  {2005})}\BibitemShut {NoStop}%
\bibitem [{\citenamefont {Schubert}\ \emph {et~al.}(2004)\citenamefont
  {Schubert}, \citenamefont {Werner}, \citenamefont {Zakharov}, \citenamefont
  {Gerth}, \citenamefont {Kolb}, \citenamefont {Long}, \citenamefont
  {G$\ddot{\textrm{o}}$sele},\ and\ \citenamefont {Tan}}]{Schub3}%
  \BibitemOpen
  \bibfield  {author} {\bibinfo {author} {\bibfnamefont {L.}~\bibnamefont
  {Schubert}}, \bibinfo {author} {\bibfnamefont {P.}~\bibnamefont {Werner}},
  \bibinfo {author} {\bibfnamefont {N.}~\bibnamefont {Zakharov}}, \bibinfo
  {author} {\bibfnamefont {G.}~\bibnamefont {Gerth}}, \bibinfo {author}
  {\bibfnamefont {F.}~\bibnamefont {Kolb}}, \bibinfo {author} {\bibfnamefont
  {L.}~\bibnamefont {Long}}, \bibinfo {author} {\bibfnamefont {U.}~\bibnamefont
  {G$\ddot{\textrm{o}}$sele}}, \ and\ \bibinfo {author} {\bibfnamefont
  {T.}~\bibnamefont {Tan}},\ }\href@noop {} {\bibfield  {journal} {\bibinfo
  {journal} {Appl. Phys. Lett.}\ }\textbf {\bibinfo {volume} {84}},\ \bibinfo
  {pages} {4968} (\bibinfo {year} {2004})}\BibitemShut {NoStop}%
\bibitem [{\citenamefont {Seifert}\ \emph {et~al.}(2004)\citenamefont
  {Seifert}, \citenamefont {Borgstrom}, \citenamefont {Deppert}, \citenamefont
  {Dick}, \citenamefont {Johansson}, \citenamefont {Larsson}, \citenamefont
  {Martensson}, \citenamefont {Skold}, \citenamefont {Svensson}, \citenamefont
  {Wacaser}, \citenamefont {Wallenberg},\ and\ \citenamefont
  {Samuelson}}]{Seif4}%
  \BibitemOpen
  \bibfield  {author} {\bibinfo {author} {\bibfnamefont {W.}~\bibnamefont
  {Seifert}}, \bibinfo {author} {\bibfnamefont {M.}~\bibnamefont {Borgstrom}},
  \bibinfo {author} {\bibfnamefont {K.}~\bibnamefont {Deppert}}, \bibinfo
  {author} {\bibfnamefont {K.}~\bibnamefont {Dick}}, \bibinfo {author}
  {\bibfnamefont {J.}~\bibnamefont {Johansson}}, \bibinfo {author}
  {\bibfnamefont {M.}~\bibnamefont {Larsson}}, \bibinfo {author} {\bibfnamefont
  {T.}~\bibnamefont {Martensson}}, \bibinfo {author} {\bibfnamefont
  {N.}~\bibnamefont {Skold}}, \bibinfo {author} {\bibfnamefont
  {C.}~\bibnamefont {Svensson}}, \bibinfo {author} {\bibfnamefont
  {B.}~\bibnamefont {Wacaser}}, \bibinfo {author} {\bibfnamefont
  {L.}~\bibnamefont {Wallenberg}}, \ and\ \bibinfo {author} {\bibfnamefont
  {L.}~\bibnamefont {Samuelson}},\ }\href@noop {} {\bibfield  {journal}
  {\bibinfo  {journal} {J.Cryst. Growth}\ }\textbf {\bibinfo {volume} {272}},\
  \bibinfo {pages} {211} (\bibinfo {year} {2004})}\BibitemShut {NoStop}%
\bibitem [{\citenamefont {Dubrovskii}\ \emph {et~al.}(2005)\citenamefont
  {Dubrovskii}, \citenamefont {Cirlin}, \citenamefont {Soshnikov},
  \citenamefont {Tonkikh}, \citenamefont {Sibirev}, \citenamefont
  {Samsonenko},\ and\ \citenamefont {Ustinov}}]{Dubr5}%
  \BibitemOpen
  \bibfield  {author} {\bibinfo {author} {\bibfnamefont {V.}~\bibnamefont
  {Dubrovskii}}, \bibinfo {author} {\bibfnamefont {G.}~\bibnamefont {Cirlin}},
  \bibinfo {author} {\bibfnamefont {I.}~\bibnamefont {Soshnikov}}, \bibinfo
  {author} {\bibfnamefont {A.}~\bibnamefont {Tonkikh}}, \bibinfo {author}
  {\bibfnamefont {N.}~\bibnamefont {Sibirev}}, \bibinfo {author} {\bibfnamefont
  {Y.}~\bibnamefont {Samsonenko}}, \ and\ \bibinfo {author} {\bibfnamefont
  {V.}~\bibnamefont {Ustinov}},\ }\href@noop {} {\bibfield  {journal} {\bibinfo
   {journal} {Phys. Rev. B}\ }\textbf {\bibinfo {volume} {71}},\ \bibinfo
  {pages} {205325} (\bibinfo {year} {2005})}\BibitemShut {NoStop}%
\bibitem [{\citenamefont {Harmand}\ \emph {et~al.}(2005)\citenamefont
  {Harmand}, \citenamefont {Patriarche}, \citenamefont {P\'{e}r\'{e}-Laperne},
  \citenamefont {M.-N.M\'{e}rat-Combes}, \citenamefont {Travers},\ and\
  \citenamefont {Glas}}]{Harm6}%
  \BibitemOpen
  \bibfield  {author} {\bibinfo {author} {\bibfnamefont {J.~C.}\ \bibnamefont
  {Harmand}}, \bibinfo {author} {\bibfnamefont {G.}~\bibnamefont {Patriarche}},
  \bibinfo {author} {\bibfnamefont {N.}~\bibnamefont {P\'{e}r\'{e}-Laperne}},
  \bibinfo {author} {\bibnamefont {M.-N.M\'{e}rat-Combes}}, \bibinfo {author}
  {\bibfnamefont {L.}~\bibnamefont {Travers}}, \ and\ \bibinfo {author}
  {\bibfnamefont {F.}~\bibnamefont {Glas}},\ }\href@noop {} {\bibfield
  {journal} {\bibinfo  {journal} {Appl. Phys. Lett.}\ }\textbf {\bibinfo
  {volume} {87}},\ \bibinfo {pages} {203101} (\bibinfo {year}
  {2005})}\BibitemShut {NoStop}%
\bibitem [{\citenamefont {Plante}\ and\ \citenamefont
  {LaPierre}(2006)}]{Plan7}%
  \BibitemOpen
  \bibfield  {author} {\bibinfo {author} {\bibfnamefont {M.~C.}\ \bibnamefont
  {Plante}}\ and\ \bibinfo {author} {\bibfnamefont {R.~R.}\ \bibnamefont
  {LaPierre}},\ }\href@noop {} {\bibfield  {journal} {\bibinfo  {journal} {J.
  Cryst. Growth}\ }\textbf {\bibinfo {volume} {286}},\ \bibinfo {pages} {394}
  (\bibinfo {year} {2006})}\BibitemShut {NoStop}%
\bibitem [{\citenamefont {Dubrovskii}\ \emph {et~al.}(2006)\citenamefont
  {Dubrovskii}, \citenamefont {Sibirev}, \citenamefont {Cirlin}, \citenamefont
  {Harmand},\ and\ \citenamefont {Ustinov}}]{Dubr8}%
  \BibitemOpen
  \bibfield  {author} {\bibinfo {author} {\bibfnamefont {V.}~\bibnamefont
  {Dubrovskii}}, \bibinfo {author} {\bibfnamefont {N.}~\bibnamefont {Sibirev}},
  \bibinfo {author} {\bibfnamefont {G.}~\bibnamefont {Cirlin}}, \bibinfo
  {author} {\bibfnamefont {J.}~\bibnamefont {Harmand}}, \ and\ \bibinfo
  {author} {\bibfnamefont {V.}~\bibnamefont {Ustinov}},\ }\href@noop {}
  {\bibfield  {journal} {\bibinfo  {journal} {Phys. Rev. E}\ }\textbf {\bibinfo
  {volume} {73}},\ \bibinfo {pages} {021603} (\bibinfo {year}
  {2006})}\BibitemShut {NoStop}%
\bibitem [{\citenamefont {Glas}\ and\ \citenamefont {Harmand}(2006)}]{Glas9}%
  \BibitemOpen
  \bibfield  {author} {\bibinfo {author} {\bibfnamefont {F.}~\bibnamefont
  {Glas}}\ and\ \bibinfo {author} {\bibfnamefont {J.~C.}\ \bibnamefont
  {Harmand}},\ }\href@noop {} {\bibfield  {journal} {\bibinfo  {journal} {Phys.
  Rev. B}\ }\textbf {\bibinfo {volume} {73}},\ \bibinfo {pages} {155320}
  (\bibinfo {year} {2006})}\BibitemShut {NoStop}%
\bibitem [{\citenamefont {Persson}\ \emph {et~al.}(2009)\citenamefont
  {Persson}, \citenamefont {Fr$\ddot{\textrm{o}}$berg}, \citenamefont
  {Jeppesen}, \citenamefont {Bj$\ddot{\textrm{o}}$rk},\ and\ \citenamefont
  {L.Samuelson}}]{Pers10}%
  \BibitemOpen
  \bibfield  {author} {\bibinfo {author} {\bibfnamefont {A.}~\bibnamefont
  {Persson}}, \bibinfo {author} {\bibfnamefont {L.}~\bibnamefont
  {Fr$\ddot{\textrm{o}}$berg}}, \bibinfo {author} {\bibfnamefont
  {S.}~\bibnamefont {Jeppesen}}, \bibinfo {author} {\bibfnamefont
  {M.}~\bibnamefont {Bj$\ddot{\textrm{o}}$rk}}, \ and\ \bibinfo {author}
  {\bibnamefont {L.Samuelson}},\ }\href@noop {} {\bibfield  {journal} {\bibinfo
   {journal} {J. Cryst. Growth}\ }\textbf {\bibinfo {volume} {311}},\ \bibinfo
  {pages} {2123} (\bibinfo {year} {2009})}\BibitemShut {NoStop}%
\bibitem [{\citenamefont {Gudiksen}\ \emph {et~al.}(2002)\citenamefont
  {Gudiksen}, \citenamefont {Lauhon}, \citenamefont {J.Wang}, \citenamefont
  {Smith},\ and\ \citenamefont {Lieber}}]{Gud11}%
  \BibitemOpen
  \bibfield  {author} {\bibinfo {author} {\bibfnamefont {M.}~\bibnamefont
  {Gudiksen}}, \bibinfo {author} {\bibfnamefont {L.}~\bibnamefont {Lauhon}},
  \bibinfo {author} {\bibnamefont {J.Wang}}, \bibinfo {author} {\bibfnamefont
  {D.}~\bibnamefont {Smith}}, \ and\ \bibinfo {author} {\bibfnamefont
  {C.}~\bibnamefont {Lieber}},\ }\href@noop {} {\bibfield  {journal} {\bibinfo
  {journal} {Nature}\ }\textbf {\bibinfo {volume} {415}},\ \bibinfo {pages}
  {617} (\bibinfo {year} {2002})}\BibitemShut {NoStop}%
\bibitem [{\citenamefont {Bj$\ddot{\textrm{o}}$rk}\ \emph
  {et~al.}(2002)\citenamefont {Bj$\ddot{\textrm{o}}$rk}, \citenamefont
  {Ohlsson}, \citenamefont {Sass}, \citenamefont {Persson}, \citenamefont
  {Thelander}, \citenamefont {Magnusson}, \citenamefont {Deppert},
  \citenamefont {Wallenberg},\ and\ \citenamefont {Samuelson}}]{Bj12}%
  \BibitemOpen
  \bibfield  {author} {\bibinfo {author} {\bibfnamefont {M.}~\bibnamefont
  {Bj$\ddot{\textrm{o}}$rk}}, \bibinfo {author} {\bibfnamefont
  {B.}~\bibnamefont {Ohlsson}}, \bibinfo {author} {\bibfnamefont
  {T.}~\bibnamefont {Sass}}, \bibinfo {author} {\bibfnamefont {A.}~\bibnamefont
  {Persson}}, \bibinfo {author} {\bibfnamefont {C.}~\bibnamefont {Thelander}},
  \bibinfo {author} {\bibfnamefont {M.~H.}\ \bibnamefont {Magnusson}}, \bibinfo
  {author} {\bibfnamefont {K.}~\bibnamefont {Deppert}}, \bibinfo {author}
  {\bibfnamefont {L.~R.}\ \bibnamefont {Wallenberg}}, \ and\ \bibinfo {author}
  {\bibfnamefont {L.}~\bibnamefont {Samuelson}},\ }\href@noop {} {\bibfield
  {journal} {\bibinfo  {journal} {Appl. Phys. Lett.}\ }\textbf {\bibinfo
  {volume} {80}},\ \bibinfo {pages} {1058} (\bibinfo {year}
  {2002})}\BibitemShut {NoStop}%
\bibitem [{\citenamefont {Cui}\ and\ \citenamefont {Lieber}(2000)}]{Cui13}%
  \BibitemOpen
  \bibfield  {author} {\bibinfo {author} {\bibfnamefont {Y.}~\bibnamefont
  {Cui}}\ and\ \bibinfo {author} {\bibfnamefont {C.~M.}\ \bibnamefont
  {Lieber}},\ }\href@noop {} {\bibfield  {journal} {\bibinfo  {journal}
  {Science}\ }\textbf {\bibinfo {volume} {91}},\ \bibinfo {pages} {851}
  (\bibinfo {year} {2000})}\BibitemShut {NoStop}%
\bibitem [{\citenamefont {Patolsky}\ \emph {et~al.}(2004)\citenamefont
  {Patolsky}, \citenamefont {Zheng}, \citenamefont {Hayden}, \citenamefont
  {Lakadamyali}, \citenamefont {X.Zhuang},\ and\ \citenamefont
  {Lieber}}]{Pat14}%
  \BibitemOpen
  \bibfield  {author} {\bibinfo {author} {\bibfnamefont {F.}~\bibnamefont
  {Patolsky}}, \bibinfo {author} {\bibfnamefont {G.}~\bibnamefont {Zheng}},
  \bibinfo {author} {\bibfnamefont {O.}~\bibnamefont {Hayden}}, \bibinfo
  {author} {\bibfnamefont {M.}~\bibnamefont {Lakadamyali}}, \bibinfo {author}
  {\bibnamefont {X.Zhuang}}, \ and\ \bibinfo {author} {\bibfnamefont {C.~M.}\
  \bibnamefont {Lieber}},\ }\href@noop {} {\bibfield  {journal} {\bibinfo
  {journal} {Proc. Natl. Acad. Sci. U.S.A.}\ }\textbf {\bibinfo {volume}
  {101}},\ \bibinfo {pages} {14017} (\bibinfo {year} {2004})}\BibitemShut
  {NoStop}%
\bibitem [{\citenamefont {Kuroda}\ \emph {et~al.}(2007)\citenamefont {Kuroda},
  \citenamefont {Kuroda}, \citenamefont {Sakoda}, \citenamefont {Kido},\ and\
  \citenamefont {Koguchi}}]{Kur18}%
  \BibitemOpen
  \bibfield  {author} {\bibinfo {author} {\bibfnamefont {K.}~\bibnamefont
  {Kuroda}}, \bibinfo {author} {\bibfnamefont {T.}~\bibnamefont {Kuroda}},
  \bibinfo {author} {\bibfnamefont {K.}~\bibnamefont {Sakoda}}, \bibinfo
  {author} {\bibfnamefont {G.}~\bibnamefont {Kido}}, \ and\ \bibinfo {author}
  {\bibfnamefont {N.}~\bibnamefont {Koguchi}},\ }\href@noop {} {\bibfield
  {journal} {\bibinfo  {journal} {J. Lumin.}\ }\textbf {\bibinfo {volume}
  {122{}-123}},\ \bibinfo {pages} {789} (\bibinfo {year} {2007})}\BibitemShut
  {NoStop}%
\bibitem [{\citenamefont {Shtrikman}\ \emph {et~al.}(2009)\citenamefont
  {Shtrikman}, \citenamefont {Popovitz-Biro}, \citenamefont {Kretinin},\ and\
  \citenamefont {Heiblum}}]{Sht15}%
  \BibitemOpen
  \bibfield  {author} {\bibinfo {author} {\bibfnamefont {H.}~\bibnamefont
  {Shtrikman}}, \bibinfo {author} {\bibfnamefont {R.}~\bibnamefont
  {Popovitz-Biro}}, \bibinfo {author} {\bibfnamefont {A.}~\bibnamefont
  {Kretinin}}, \ and\ \bibinfo {author} {\bibfnamefont {M.}~\bibnamefont
  {Heiblum}},\ }\href@noop {} {\bibfield  {journal} {\bibinfo  {journal} {Nano
  Lett.}\ }\textbf {\bibinfo {volume} {9}},\ \bibinfo {pages} {215} (\bibinfo
  {year} {2009})}\BibitemShut {NoStop}%
\bibitem [{\citenamefont {Dubrovskii}\ \emph {et~al.}(2009)\citenamefont
  {Dubrovskii}, \citenamefont {Sibirev}, \citenamefont {Cirlin}, \citenamefont
  {Bouravleuv}, \citenamefont {Samsonenko}, \citenamefont {Dheeraj},
  \citenamefont {Zhou}, \citenamefont {Sartel}, \citenamefont {Harmand},
  \citenamefont {Patriarche},\ and\ \citenamefont {Glas}}]{Dub16}%
  \BibitemOpen
  \bibfield  {author} {\bibinfo {author} {\bibfnamefont {V.}~\bibnamefont
  {Dubrovskii}}, \bibinfo {author} {\bibfnamefont {N.}~\bibnamefont {Sibirev}},
  \bibinfo {author} {\bibfnamefont {G.}~\bibnamefont {Cirlin}}, \bibinfo
  {author} {\bibfnamefont {A.}~\bibnamefont {Bouravleuv}}, \bibinfo {author}
  {\bibfnamefont {Y.}~\bibnamefont {Samsonenko}}, \bibinfo {author}
  {\bibfnamefont {D.}~\bibnamefont {Dheeraj}}, \bibinfo {author} {\bibfnamefont
  {H.}~\bibnamefont {Zhou}}, \bibinfo {author} {\bibfnamefont {C.}~\bibnamefont
  {Sartel}}, \bibinfo {author} {\bibfnamefont {J.}~\bibnamefont {Harmand}},
  \bibinfo {author} {\bibfnamefont {G.}~\bibnamefont {Patriarche}}, \ and\
  \bibinfo {author} {\bibfnamefont {F.}~\bibnamefont {Glas}},\ }\href@noop {}
  {\bibfield  {journal} {\bibinfo  {journal} {Phys. Rev. B}\ }\textbf {\bibinfo
  {volume} {80}},\ \bibinfo {pages} {205305} (\bibinfo {year}
  {2009})}\BibitemShut {NoStop}%
\bibitem [{\citenamefont {Spirkoska}\ \emph {et~al.}(2009)\citenamefont
  {Spirkoska}, \citenamefont {Arbiol}, \citenamefont {Gustafsson},
  \citenamefont {Conesa-Boj}, \citenamefont {Glas}, \citenamefont {Zardo},
  \citenamefont {Heigoldt}, \citenamefont {Gass}, \citenamefont {Bleloch},
  \citenamefont {Estrade}, \citenamefont {Kaniber}, \citenamefont {Rossler},
  \citenamefont {Peiro}, \citenamefont {Morante}, \citenamefont {Abstreiter},
  \citenamefont {Samuelson},\ and\ \citenamefont {i~Morral}}]{Spir17}%
  \BibitemOpen
  \bibfield  {author} {\bibinfo {author} {\bibfnamefont {D.}~\bibnamefont
  {Spirkoska}}, \bibinfo {author} {\bibfnamefont {J.}~\bibnamefont {Arbiol}},
  \bibinfo {author} {\bibfnamefont {A.}~\bibnamefont {Gustafsson}}, \bibinfo
  {author} {\bibfnamefont {S.}~\bibnamefont {Conesa-Boj}}, \bibinfo {author}
  {\bibfnamefont {F.}~\bibnamefont {Glas}}, \bibinfo {author} {\bibfnamefont
  {I.}~\bibnamefont {Zardo}}, \bibinfo {author} {\bibfnamefont
  {M.}~\bibnamefont {Heigoldt}}, \bibinfo {author} {\bibfnamefont {M.~H.}\
  \bibnamefont {Gass}}, \bibinfo {author} {\bibfnamefont {A.~L.}\ \bibnamefont
  {Bleloch}}, \bibinfo {author} {\bibfnamefont {S.}~\bibnamefont {Estrade}},
  \bibinfo {author} {\bibfnamefont {M.}~\bibnamefont {Kaniber}}, \bibinfo
  {author} {\bibfnamefont {J.}~\bibnamefont {Rossler}}, \bibinfo {author}
  {\bibfnamefont {F.}~\bibnamefont {Peiro}}, \bibinfo {author} {\bibfnamefont
  {J.~R.}\ \bibnamefont {Morante}}, \bibinfo {author} {\bibfnamefont
  {G.}~\bibnamefont {Abstreiter}}, \bibinfo {author} {\bibfnamefont
  {L.}~\bibnamefont {Samuelson}}, \ and\ \bibinfo {author} {\bibfnamefont
  {A.~F.}\ \bibnamefont {i~Morral}},\ }\href@noop {} {\bibfield  {journal}
  {\bibinfo  {journal} {Phys. Rev. B}\ }\textbf {\bibinfo {volume} {80}},\
  \bibinfo {pages} {245325} (\bibinfo {year} {2009})}\BibitemShut {NoStop}%
\end{thebibliography}

\end{document}